\documentclass[12pt]{JHEP3}
\usepackage{amsfonts}
\usepackage{amssymb}
\usepackage{lscape}
\usepackage{cite}
\usepackage{epsfig}
\usepackage{multirow}
\usepackage[all]{xy}
\usepackage{amsmath}
\usepackage{longtable}
\author{}

\newcommand{\drawsquare}[2]{\hbox{%
\rule{#2pt}{#1pt}\hskip-#2pt
\rule{#1pt}{#2pt}\hskip-#1pt
\rule[#1pt]{#1pt}{#2pt}}\rule[#1pt]{#2pt}{#2pt}\hskip-#2pt
\rule{#2pt}{#1pt}}
\newcommand{\fund}{\raisebox{-.5pt}{\drawsquare{6.5}{0.4}}}

\newcommand{\be}{\begin{equation}}
\newcommand{\ee}{\end{equation}}
\newcommand{\ba}{\begin{array}}
\newcommand{\ea}{\end{array}}
\newcommand{\bea}{\begin{eqnarray}}
\newcommand{\eea}{\end{eqnarray}}

\newcommand{\ov}{\overline}
\def\IR{\relax{\rm I\kern-.18em R}}

\def\IP{\relax{\rm I\kern-.18em P}}
\def\inbar{\vrule height1.5ex width.4pt depth0pt}
\def\IC{\relax\,\hbox{$\inbar\kern-.3em{\rm C}$}}

\def\K3{{\bf K3}}

\def\ov{\overline}

\def\n2d{\cN_{V^*}^{\otimes 2}}

\def\IC{\mathbb{C}}

\def\IR{\mathbb{R}}

\def\IP{\mathbb{P}}

\def\cN{{\mathcal N}}

\def\nn{\nonumber}

\title{Discrete symmetries from hidden sectors}
\author{
Pascal Anastasopoulos$^{1}$\footnote{pascal@hep.itp.tuwien.ac.at},
Robert Richter$^{2}$\footnote{robert.richter@desy.de} and
A. N. Schellekens$^{3,4,5}$\footnote{t58@nikhef.nl}~
\\
$^1$ Technische Univ. Wien Inst. f\"ur Theoretische Physik, A-1040 Vienna, Austria\\
$^2$ II. Institut f\"ur Theoretische Physik, Hamburg University, Germany\\
$^3$ NIKHEF, Science Park 105, 1098 XG Amsterdam, The Netherlands\\
$^4$ IMAPP, Radboud Universiteit Nijmegen, The Netherlands\\
$^5$ Instituto de F\'\i sica Fundamental, CSIC, Madrid, Spain 
}
\date{}

\maketitle
\abstract{We study the presence of abelian discrete symmetries in globally consistent orientifold compactifications based on rational conformal field theory. We extend previous work \cite{Ibanez:2012wg} by allowing the discrete symmetries to be a linear combination of $U(1)$ gauge factors of the visible as well as the hidden sector. This more general ansatz significantly increases the probability of finding a discrete symmetry in the low energy effective action. Applied to globally consistent MSSM-like Gepner constructions we find multiple models that allow for matter parity or Baryon triality. 
}
\preprint{TUW-15-03\\ 
ZMP-HH/15-2\\
Nikhef/2015-004}
 \clearpage

\begin{document}

\section{Introduction}

Discrete symmetries play a prominent role in various new theoretical ideas in high energy physics. They are often times imposed to ensure stability of the proton in extensions of the Standard Model (SM) \cite{Ibanez:1991pr} as well as to explain the observed mass hierarchies of the SM fermions \cite{Altarelli:2010gt}. Furthermore they may play a crucial role in cosmology, e.g. one often invokes a discrete symmetry to ensure the absence of higher dimensional operators whose presence would otherwise spoil the inflaton potential (see, e.g. \cite{Baumann:2009ds}).

In type II string compactifications abelian discrete symmetries generically arise as discrete subgroups of anomalous $U(1)$ gauge factors. A typical D-brane compactification exhibits multiple $U(1)$'s which typically appear anomalous. Their anomalies are cancelled by the Green-Schwarz mechanism \cite{Sagnotti:1992qw,Ibanez:1998qp,Bianchi:2000de,Aldazabal:2000dg,Antoniadis:2002cs} which makes the $U(1)$'s massive via St\"uckelberg couplings. 
In the low energy effective theory the massive $U(1)$'s survive as global symmetries that are preserved by all perturbative quantities. However, non-perturbative effects, so called D-instantons, can break those global symmetries thus inducing desired couplings that are absent perturbatively \cite{Blumenhagen:2006xt,Ibanez:2006da,Blumenhagen:2009qh}. 
A discrete symmetry $\mathbb{Z}_N$, preserved by all perturbative and non-perturbative quantities, is then given by a linear combination of the anomalous $U(1)$ gauge factors, under which all possible D-instanton configurations in the global compactification will be uncharged~\cite{BerasaluceGonzalez:2011wy} (see also \cite{Ibanez:2012wg,Anastasopoulos:2012zu,Honecker:2013hda,Honecker:2013kda, Honecker:2015ela})\footnote{Discrete flavor symmetries in D-brane compactifications that originate from isometries of the compactification manifold have been discussed in \cite{Abe:2009vi,BerasaluceGonzalez:2012vb,Marchesano:2013ega,Hamada:2014hpa}. For a discussion on discrete symmetries in heterotic compactifications, see \cite{Kobayashi:2006wq,Forste:2010pf,Nilles:2012cy} and in the context of F-theory compactifications, see \cite{Braun:2014oya,Morrison:2014era,Anderson:2014yva,Karozas:2014aha,Klevers:2014bqa,Garcia-Etxebarria:2014qua,Mayrhofer:2014haa,Mayrhofer:2014laa,Braun:2014qka,Leontaris:2015yva}.}.

The aim of this work is to extend the analysis performed in \cite{Ibanez:2012wg}. There the authors discussed the presence of abelian discrete symmetries in globally consistent Gepner constructions that exhibit a visible sector that contains the SM or extensions of it. Their systematic search revealed that discrete symmetries that arise solely from $U(1)$ gauge factors of the visible sector are very rarely realized. Here we generalize this analysis by allowing the discrete symmetries to be linear combinations of visible as well as hidden $U(1)$ gauge factors. With this generalization we find the  probability for the presence of abelian discrete symmetries is significantly increased. Moreover, we find explicit global MSSM Gepner constructions that exhibit matter parity or Baryon triality and thus can explain the stability of the proton. 

The essential difference between the analysis of \cite{Ibanez:2012wg} and the present paper is that the order of two computations is interchanged. The first step
in \cite{Ibanez:2012wg} was to determine the discrete symmetries of local standard model realizations. Then a search was performed for additional ``hidden sector" branes
to cancel the dilaton tadpoles. Instead, we search {\it first} for tadpole cancelling hidden sectors, and {\it then} we determine the discrete symmetries. In this manner, {\it all} $U(1)$ symmetries of the full global configuration can contribute to discrete symmetries, whereas in the procedure followed in \cite{Ibanez:2012wg} only the $U(1)$'s of the Standard Model
sector contribute. Therefore our results contain those of \cite{Ibanez:2012wg} as a special case.

The additional discrete symmetries we find may act entirely within the hidden sector, or on both observable and hidden sector particles. The former case is not of much interest,
because it cannot  constrain observable sector couplings. 
In the latter case the additional discrete symmetries may give rise to novel restrictions on couplings between observable sector particles.  

Note that there may also exist mixed observable-hidden massless $U(1)$'s not seen in the local model. To illustrate that, consider the class of most interest, the 
$U(3)\times Sp(2) \times U(1)\times U(1)$  Madrid-type configuration \cite{Ibanez:2001nd}. Generically, this class has two massless $U(1)$'s, namely the Standard Model gauge symmetry $Y$, plus
a gauged $B-L$.  The presence of the latter makes further analysis of these models uninteresting. Not only does the massless $B-L$ disagree with observation, and require
an additional (Higgs) mechanism to give mass to the $B-L$ gauge boson, but furthermore the massless $B-L$ already forbids all dimension four $B$ and $L$
violating processes. It makes little sense to look for discrete symmetries to forbid the less important dimension five operators (not forbidden by $B-L$) without implementing a $B-L$ breaking mechanism first. However, there is a small subclass where the $B-L$ gauge boson already gets a mass from axion mixing in string theory. We will focus on this class  not only for the reason stated above, but also for practical reasons: our approach requires a systematic search of tadpole solutions, and this becomes extremely time-consuming for the much larger, but much less interesting class with unbroken $B-L$. 

Even if we start with a class with a broken $B-L$ symmetry, mixing with hidden sector $U(1)$'s may lead to massless $U(1)$'s that act on the Standard Model particles. Phenomenologically this is as disastrous as a massless $B-L$, and hence in the spirit of our statements above all such tadpole cancellation solutions must be rejected. Usually this just removes a subset of all tadpole solutions. Within that subset, it may happen that a mixed observable-hidden $U(1)$ is not completely broken, but is broken to a discrete symmetry. These are the cases of interest.

\section{Discrete symmetries in Gepner constructions}

Gepner orientifold constructions have been proven to be a fruitful framework for semi-realistic string model building \cite{Dijkstra:2004ym,Dijkstra:2004cc,Anastasopoulos:2006da}  \footnote{For other work on string spectra in Gepner orientifolds, see \cite{Blumenhagen:2003su,Blumenhagen:2004hd,Blumenhagen:2004qu,Blumenhagen:2004cg,Aldazabal:2003ub,Govindarajan:2003vp,Aldazabal:2004by,Brunner:2004zd}.}. The basic building blocks are boundary and cross cap states in a rational conformal field theory, which provide the analogue of D-branes and orientifold planes, the basic ingredients of D-brane constructions. In full analogy to D-brane compactifications one can determine the gauge symmetry, specifically the presence of $U(1)$'s in the low energy effective action, compute the massless spectrum, as well as check global consistency. To check the latter we verify that all tadpoles cancel and that all K-theory charges cancel as well \cite{Uranga:2000xp}.

The only class of rational conformal field theories where the relevant quantities can be computed are tensor products of $N=2$ minimal models, originally considered for
heterotic string model building by Gepner \cite{Gepner:1987qi}. The formalism for describing boundary and cross cap states in these closed string CFT's was developed in
a large number of papers, mainly during the last decade of the last century. See \cite{Angelantonj:2002ct} for a review and further references. The main feature that makes $N=2$ minimal models such a rich area for model building is its large simple current symmetry group. The general formalism for simple current modifications of boundary and cross cap states was presented in its definitive form in \cite{Fuchs:2000cm}.



The approach to model building generally used in RCFT orientifold constructions is to search first for a set of three or four boundary states (``D-branes")\footnote{Complex brane stacks are counted as one.}  whose
Chan-Paton gauge group contains the Standard Model, and whose chiral intersections correspond to three families of quarks and leptons. We will refer to such a 
configuration as a \emph{local SM Gepner configuration}.
Then, in a second step,
a search is done for additional branes that cancel the tadpoles (in a very small number of cases all  tadpoles cancel among the Standard Model branes themselves).
This then yields a \emph{global SM Gepner configuration}.
In \cite{Dijkstra:2004ym,Dijkstra:2004cc} local configurations of the so called ``Madrid-type" (and mild generalisations thereof) were searched for, and a very deep search for tadpole solutions was performed.  The main defining features of a Madrid type configuration is that all quarks and leptons originate from bi-fundamentals, and that there is
a well-defined global baryon and lepton number symmetry expressed in terms of brane charges.

In \cite{Anastasopoulos:2006da} the definition of a local model was generalized to arbitrary configurations with at most four branes, limited only by the
requirement that the chiral spectrum agree with the Standard Model if the full Chan-Paton gauge group 
is group-theoretically reduced to its subgroup $SU(3)\times SU(2)\times U(1)$.  The full Chan-Paton group is the one directly obtained from the brane multiplicities, without
taking into account mixing of $U(1)$ gauge bosons with axions. It must contain $SU(3)\times SU(2)\times U(1)$, but the existence of a  physical mechanism to realise the
reduction is not part of the requirements.


The first step of this search yielded  19345 chirally distinct  local SM Gepner configurations. Two local SM Gepner configuration are distinct if they have a different 
full Chan-Paton gauge group or different chiral spectra with respect to that group. Furthermore two local SM Gepner configurations are distinct if the number of massless
$U(1)$'s after axion mixing is different. We will refer to these 19345 spectra as \emph{ADKS classes} henceforth.
The second step in \cite{Anastasopoulos:2006da} was a search for tadpole solutions with a limit on the number of hidden branes (the limit was determined by CPU-time constraints). 
If this search was successful, no further tadpole solution searches were done for the same ADKS class. The only goal was to establish that there does not exist any obstruction to
realizing a certain ADKS class globally, and a single example establishes that. The authors found 
that a global configuration exists for around 10 percent of the 19345 ADKS classes. 
Note that this does not imply non-existence of global realizations in the other cases. A database of the 19345 ADKS classes with all their explicit brane realizations (specified by means of minimal model combinations, the choice of modular invariant partition functions and orientifold projections, and the choice of three or four boundary state labels)
is available
\footnote{See www.nikhef.nl/$\sim$t58/Site/String\_Spectra.html. In the database all 19345 spectra are labelled by an integer, which provides a convenient way to refer to a class. The same labels were used in other papers, in particular \cite{Ibanez:2012wg}. We will refer to this as the \emph{ADKS-label}.}, and is the starting point of our analysis. Since we need global SM Gepner configurations, we will not consider any  ADKS class where no tadpole solution was found in
\cite{Anastasopoulos:2006da}, because if they exist they would be very hard to find.


Since 2006 lot of progress has been made by investigating in more detail the phenomenology of those constructions \cite{Ibanez:2007rs, Kiritsis:2008ry,Kiritsis:2009sf,Anastasopoulos:2010ca,Anastasopoulos:2010hu }. In particular a lot of effort has been dedicated to analyze the superpotential couplings of those models, by also taking D-instanton effects \cite{Blumenhagen:2006xt,Ibanez:2006da,Blumenhagen:2009qh} into account.

In a recent work \cite{Ibanez:2012wg} the authors discussed the presence of abelian discrete symmetries in globally consistent Gepner constructions. They give an explicit procedure that allows to systematically search for a discrete symmetry in a given Gepner construction. More precisely, they provide a method to obtain an integral basis for the couplings of axions to the $U(1)$ gauge factors. Given this integral basis they investigated a subset of the 19345 ADKS classes found in \cite{Anastasopoulos:2006da} 
with respect to the presence of abelian discrete symmetries. The considered subclass consists of 4 D-brane stacks yielding the D-brane gauge symmetries
\begin{align}
U(3)_a \times U(2)_b \times U(1)_c \times U(1)_d \hspace{8mm} \text{or} \hspace{8mm} U(3)_a \times Sp(2)_b \times U(1)_c \times U(1)_d \,\,,
\end{align}  
where the subscript $a$ to $d$ enumerate the different D-brane stacks, $a$ denoting the color and $b$ the weak force D-brane stacks. The hypercharge of the considered cases is chosen to be given by the Madrid embedding, namely by the linear combination

\begin{align}
U(1)_Y= \frac{1}{6}U(1)_a + \frac{1}{2} U(1)_c - \frac{1}{2} U(1)_d\,\,.
\label{eq hypercharge madrid}
 \end{align}  
Finally, the authors required that all chiral SM fields are realized as bi-fundamentals under the D-brane gauge symmetries. In total the subclass of all considered local Gepner construction consists of 24 different local D-brane configurations.

Within this class of 24 local SM Gepner configurations the authors found only very few examples that exhibit a discrete symmetry. Moreover, when searching for global embeddings of these local configurations that exhibit a discrete symmetries the authors found only examples that exhibit a $\mathbb{Z}_2$ that can be interpreted as matter parity or models that exhibit a $\mathbb{Z}_3$ but on top of that also an additional $U(1)$ gauge symmetry. As explained in the introduction, this is an undesirable feature one has to address before a meaningful discussion of discrete symmetries can be made.

Let us stress once again that this procedure searches for potential discrete symmetries that are given by a linear combination of the $U(1)$ factors of the SM D-branes. 
In this work we want to generalize this ansatz by allowing the discrete symmetries to be a linear combination of $U(1)$ gauge factors originating from the visible but also hidden sector. In order to perform such a analysis one has to significantly change the search procedure. In contrast to the search performed in \cite{Ibanez:2012wg} here we first generate  for a given local Gepner SM configuration a large number of tadpole free models. Within these globally consistent models we analyze the axion couplings and determine whether they exhibit a discrete symmetry. Eventually we will investigate the consequences of those discrete symmetries. 

In order to keep the search feasible we constrain ourselves to allow the hidden sector to contain at most 5 D-brane stacks. Moreover, we focus on one specific ADKS class, with ADKS-label 7506.  It exhibits the D-brane gauge symmetry
\begin{align} 
U(3)_a \times Sp(2)_b \times U(1)_c \times U(1)_d
\end{align}
with the hypercharge taking the form given by eq. \eqref{eq hypercharge madrid} and no additional massless $U(1)$ in the SM sector. In table \ref{table spectrum Nr. 7506} we display the transformation behavior of the chiral SM spectrum transforms under the D-brane gauge symmetries. Note that all desired Yukawa couplings $Q_L H_u U_R$, $Q_L H_d D_R$,
$L H_d E_R$, and $L H_u N_R$ are allowed on the perturbative level, i.e. they are neutral under the all the $U(1)$ gauge factors of the individual D-brane stacks. On the other hand 
the Majorana mass term for the right-handed neutrinos $N_R N_R$, as well as the dangerous dimension 4 and 5 proton decay operators $U_R D_R D_R$, $L L E_R$, $Q_L Q_L Q_L L$ and $U_R U_R D_R E_R$ are perturbatively forbidden. However, the latter can be induced non-perturbatively by D-instantons.  Even though being exponentially suppressed, their presence poses severe problems for the stability of the proton. Therefore it is desirable to have a discrete symmetry that forbids such proton decay operators, while still allowing for a Majorana mass term for the right-handed neutrinos to account for the small neutrino masses. 

\begin{table}[h]
\center
\hspace{0.00cm}
\scalebox{0.85}{
\begin{tabular}{|c|c|c|c|c|c|c|c|c|}\hline
Matter fields &$Q_L$&$D_R$&$U_R$&$L$&$E_R$&$N_R$&$H_u$&$H_d$\\ \hline 
Transformation & $(\fund_a,\fund_b)$
& $(\ov\fund_{a},\fund_c)$ 
& $(\ov\fund_{a},\ov\fund_{c})$ 
& $(\fund_b,\fund_d)$ 
& $(\fund_c,\ov\fund_{d})$ 
& $(\ov\fund_{c},\ov\fund_{d})$ 
& $(\fund_b,\fund_c)$ 
& $(\fund_b,\ov \fund_c)$ 
\\
\hline
Multiplicity  &3&3&3&3&3&3&n&n\\ \hline		
\end{tabular}}
\caption{\small Chiral SM spectrum for the local D-brane configuration Nr. 7506.
\label{table spectrum Nr. 7506}}
\end{table}

Let us briefly discuss what discrete symmetries we expect to find\footnote{This discussion will focus solely on the visible sector. At this stage nothing can be said about the hidden sector.}. 
It can be shown \cite{Ibanez:1991pr,Ibanez:1991hv} that for any family independent discrete gauge symmetry 
$\mathbf{Z}_N$ in the MSSM with three right-handed neutrinos the corresponding generator $g_N$ can be written as
 \begin{align}
g_N = A^{n}_N  \times  L^p_N \times R^m_N\,\,,
\end{align}
where   $A_N$, $L_N$ and $R_N$ are mutually commuting generators and the exponents run over $m, n, p = 0, 1, ... N-1$. 
In table \ref{table charges under three Z_N} we display the charges of the MSSM matter fields under the respective generators, where one imposes that all desired Yukawa couplings are allowed by the discrete symmetry. 

\begin{table}[h]
\centering
\begin{tabular}{|c|c|c|c|c|c|c|c|c|}
\hline
    & $Q_L$ & $U_R$ & $D_R$ & $L$  & $E_R$ & $N_R$ & $H_u$ & $H_d$\\ 
\hline 
\hline
$A$ & $0$   & $0$   & $-1$  & $-1$ & $0$   & $1$   & $0$   & $1$\\  
$L$ & $0$   & $0$   & $0$   & $-1$ & $1$   & $1$   & $0$   & $0$\\
$R$ & $0$   & $-1$  & $1$   & $0$  & $1$   & $-1$  & $1$   & $-1$\\
\hline
\end{tabular}
\caption{\small The generators of family independent discrete $\mathbf{Z}_N$'s in the MSSM.}
\label{table charges under three Z_N}
\end{table}

In \cite{Dreiner:2005rd} (see, also \cite{Anastasopoulos:2012zu}) the authors classify all possible abelian discrete symmetries that satisfy the discrete gauge anomaly conditions for the MSSM. They find only $\mathbb{Z}_2$, $\mathbb{Z}_3$ and $\mathbb{Z}_6$ symmetries which meet the constraints. Their phenomenological implications are displayed in table \ref{table discrete couplings}.
\begin{table}[htb]
\scalebox{1.1}{
\begin{tabular}{|c||c||c|c|c|c||c|c|c|c||c| }
\hline
coupling         & $R_2$      & $ L_3 R_3 $ & $R_3$      & $ L_3$     &$ L_3 R^2_3 $ &$ L^2_6 R^5_6$& $R_6$      &$L^2_6 R^3_6 $&$L^2_6 R_6 $  \\
\hline
\hline
$H_u H_d$        &$\checkmark$&$\checkmark$&$\checkmark$&$\checkmark$&$\checkmark$&$\checkmark$ &$\checkmark$&$\checkmark$ &$\checkmark$\\
\hline
$L H_u$          &            &$\checkmark$&            &            &            &             &            &             &            \\
\hline
$L L E_R$        &            &$\checkmark$&            &            &            &             &            &             &            \\
\hline
$Q_L L D_R$      &            &$\checkmark$&            &            &            &             &            &             &            \\
\hline
$U_R D_R D_R$    &            &            &            &$\checkmark$&            &             &            &             &            \\
\hline
$Q_L Q_L Q_L L$  &$\checkmark$&            &$\checkmark$&            &            &             &$\checkmark$&             &            \\
\hline
\hspace{-0.5mm}$U_R U_R D_R E_R$\hspace{-0.5mm}&$\checkmark$&            &$\checkmark$&            &            &             &$\checkmark$&             &            \\
\hline
$L H_u L H_u$    &$\checkmark$&$\checkmark$&            &            &            &$\checkmark$ &            &             &            \\
\hline \hline
$N_R N_R$    &$\checkmark$&$\checkmark$&            &            &            &$\checkmark$ &            &             &            \\
\hline
\end{tabular}}
\caption{\small Allowed terms for the respective discrete gauge symmetries \cite{Dreiner:2005rd}.}
\label{table discrete couplings}
\end{table}
Particular intriguing discrete symmetries are matter parity $R_2$ and Baryon triality $L_3 R_3$, as well as the product of both proton hexality $ L^2_6 R^5_6$, that all allow the Majorana mass term for the right-handed neutrinos and at the same time forbid some (or all) dangerous dimension 4 and 5 proton decay operators. 

In the next section we discuss to what extent these discrete symmetries are realized in globally consistent MSSM Gepner constructions.   

\section{Results}

Here we present the results of the performed search. Recall that we search for all tadpole free solutions to a local Gepner construction with ADKS-label 7506, where the hidden sector consists of at most 5 hidden D-brane stacks. 
The list of models of ADKS class 7506 obtained in \cite{Anastasopoulos:2006da} has 40590 items. These lists usually have degeneracies due to interchanges of identical factors in the tensor product, conjugations in the CFT's and for various other reasons.
At present, here is no complete fundamental understanding of these degeneracies, but they are obviously a nuisance, since all search efforts are multiplied by a large
factor.  To reduce this we compared the non-chiral spectra (by definition of an ADKS class, all {\em chiral} spectra are identical) and the set of potentially allowed hidden sector branes and their properties.  We distinguish these features with a total of 37 parameters. If all these parameters are identical, we regard two models as degenerate, and consider only
one model from each degenerate set. This reduces the list from 40590 to  4911. Note that this approach differs from the one used in \cite{Ibanez:2012wg}. 
There are 40590 items on the list were first scanned for discrete symmetries. Since this is a relatively easy search, this was doable. But searching for tadpole solutions is considerably more time-consuming, and therefore we decided to remove the apparent degeneracies, at the very small risk of  removing spectra that are distinct, but not distinguishable on the basis of the 37 integers we compared. 

For each local model we allowed as hidden sector branes all brane stacks with a non-chiral intersection with the Standard Model sector. Furthermore we identified brane stacks with the same contribution to all tadpole conditions. This avoids a degeneracy in the space of tadpole solutions. For example, if a number of O-type branes all have the same coupling to all tadpoles, this results in orthogonal groups that can be split up in any possible way. We replace this set by just a single variable (the sum of all multiplicities) in the tadpole equations, so that all those identified branes will be on top of each other. This resulted in a number of tadpole loop variables ranging from 12 to 189.

The success rate of the tadpole search was almost 6\%: we found at least one tadpole solutions for 280 of the 4911 non-degenerate local models. 
Note that in \cite{Ibanez:2012wg} only one tadpole solution was considered for each local model, since this was sufficient as a proof of existence. On the other hand,
we scanned the full configuration space of the hidden sector, with a maximum of 5 stacks, 
allowing for more than one solution for each local model. In total we found 5020 tadpole free solutions, an average of about 18 per local model. If the limitation to at most five hidden sector branes stacks is lifted, this number is likely to increase drastically, but in practice this is extremely time-consuming. Note that our limit, five, is much smaller than
the total number of variables, so that each increase of this number by one increases the amount of computer time by one or two orders of magnitude.

Clearly this set is too small to draw any rigorous statistical conclusions, but it is still large enough that it may hint towards some general lessons. 

\subsection{Massless $U(1)$'s}

Before discussing the discrete symmetries let us briefly discuss the presence of massless $U(1)$'s in this set of tadpole free solutions.

The local Gepner constructions are set up in such a way that only one linear combination of the three $U(1)$ factors\footnote{Recall the local configuration Nr. 7506 is based on 4 D-brane stacks, however the weak force is realized as $Sp(2)$ gauge symmetry. Thus there are only 3 $U(1)$ gauge factors, namely the one arising from the color D-brane stack and the two originating form the single D-brane stacks.} remains massless and eventually is interpreted as hypercharge. This linear combination will remain also massless when adding a hidden D-brane sector that is required to cancel all tadpoles. However, these global completion may now allow for additional massless $U(1)$'s. In table \ref{table massless U(1)'s} 
we show the frequency of their appearances. The first two columns display how many tadpole free solutions to a specific number of massless $U(1)$'s exist, while the columns 3 to 8 expose how many unitary brane stacks there are in the global compactification, in other words, the number of $U(1)$ factors without taking into account the St\"uckelberg mechanism generated by mixing with axions.
Let us stress that the number $U(1)$ factors does not coincide with the number of D-brane stacks in a specific model, since D-brane stacks that give rise to orthogonal or symplectic gauge symmetries do not carry any $U(1)$ factors.
\begin{table}[h]
\centering
{
\begin{tabular}{|c||c||c|c|c|c|c|c|}\hline
\multirow{2}{*}{massless $U(1)$'s} &  \multirow{2}{*}{total $\#$ of solutions}  & \multicolumn{6}{|c|}{Unitary brane stacks} \\
\cline{3-8}
 & & 3 & 4 & 5 &6 & 7 & 8 \\ \hline \hline
1  &  3519 & 2267 & 784 & 201 & 130 & 60 & 77  \\ \hline
2  & 1015  & & 134 & 88 & 186 & 93 & 414    \\ \hline
3  &    404   & & & 23 & 45 & 172 & 164   \\ \hline
4  &    178  &  & & & & 20 & 158 \\ \hline
5  &   4   &  & & & &  & 4  \\ \hline
\end{tabular}}
\caption{\small Global models compared with respect to their massless $U(1)$'s.
\label{table massless U(1)'s} 
}
\end{table}

The naive expectation that the more unitary brane stacks there are  in a given model, the more likely it is to have multiple massless $U(1)$'s is met in this small sample. This fact is of importance in the enterprise of realizing the MSSM within string theory. Since additional $U(1)$'s may forbid desired couplings in the SM, such as Majorana masses for right-handed neutrinos, the above statement suggests that fewer D-brane stacks are preferred to realize just a single $U(1)$ in a global realization. 

\begin{table}[h]
\centering
{
\begin{tabular}{|c||c|c|c|}\hline
\# Unitary stacks  & \# of solutions & \# 
with just $U(1)_Y$ (visible) & ratio \\ \hline \hline
3 & 2267  &  2267 & 100\% \\ \hline
4 & 918   &   882   & 96\%  \\ \hline
5 &  312   &    285   & 91\%  \\ \hline
6 &  361   &    268   & 74\%   \\ \hline
7 &  345   &    122   &  35\% \\ \hline
8 &  817   &   204   & 25\%  \\ \hline\hline
combined &  5020   &   4028   & 80\%  \\ \hline
\end{tabular}}
\caption{\small Global models compared with respect to visible $U(1)$'s.
\label{table massless visible U(1)'s} 
}
\end{table}

Note that ADKS class 7506 has the special property that the $B-L$ photon is massive, which happens in only a few percent of such models. The corresponding class with a massless $B-L$ photon has number 2751, and
consists of 869428 local model, more than twenty times as many. 
As explained in the introduction, this feature is easily lost if a hidden sector is added. If mixing between observable and hidden $U(1)$'s leads to an additional massless $U(1)$ that couples to quarks and leptons, 
then this is equally bad as a massless $B-L$, and this makes the restriction from class 2751 to 7506 irrelevant. We might then as well have started with the full class 2751. However, the presence of an additional 
$U(1)$ is not necessarily dangerous from the MSSM string model building point of view, since it may happen that the SM fields are completely uncharged under the additional massless $U(1)$'s. 
Such a purely hidden sector $U(1)$ may still lead to subtle observable effects through kinetic mixing \cite{Holdom:1985ag}, but the consequences are not necessarily phenomenologically fatal.

In the sample of 5020 tadpole-free solutions we find 4028 solutions  that have in the visible sector only one massless $U(1)$, namely the hypercharge, whereas 3519 models out of those 4028 exhibit only one massless $U(1)$ in total, and no hidden $U(1)$'s. Hence about 80\% of all tadpole-free models are phenomenologically acceptable by the standards formulated above.
Nevertheless even taking into account the fact that one may have purely hidden $U(1)$'s the previous statement holds true. The more $U(1)$ gauge factors there are in the global completion the more likely it is that the SM string realization suffers under an additional $U(1)$, under which the SM matter fields are charged. This point is demonstrated in table \ref{table massless visible U(1)'s} which indicates that it is more likely to avoid additional massless $U(1)$'s in the visible sector for fewer $U(N)$ D-brane stacks.

On the other hand, as we will see momentarily more $U(1)$ gauge factors increase the probability to find discrete symmetries. Thus there will be a tension between having a discrete symmetry and at the same time only one massless $U(1)$ symmetry in the visible sector.  

\subsection{Discrete symmetries}

After discussing the frequency of massless $U(1)$'s in the global Gepner constructions, let us turn to the abelian discrete symmetries, the main focus of this search. We start by discussing the presence of discrete symmetries  within the sample of 5020 globally consistent, i.e. tadpole free Gepner constructions of type Nr. 7506.  From table \ref{table discrete symmetries in total} we see that about 17\% of all the models exhibit a discrete symmetry. 
We find 707 models if we furthermore require that the models exhibit a discrete symmetry under which the SM fields are charged, which is 14\% of the total number of global constructions. 
If we restrict ourselves to phenomenologically acceptable cases without spurious ``photons", we find that in 372 out of 4028 models (about 9\%) there is a discrete symmetry that acts on the Standard Model.
\begin{table}[h] \centering
\begin{tabular}{|l|c|c|c|}
\hline
total \#  of models & \#  with no $\mathbb{Z}_N$'s &  \# with  $\mathbb{Z}_N$'s  & \#  with visible $\mathbb{Z}_N$'s\\
\hline \hline
5020 (Total)\hfill   & 4189 & 831 & 707  \\ \hline
4028 (Only $Y$ visible)\hfill   & 3602 & 426 & 372  \\ \hline
 \end{tabular}\nn
\caption{\small {Discrete symmetries in globally consistent models of type Nr. 7506.}} 
\label{table discrete symmetries in total}
\end{table}

Let us now compare our results to those of \cite{Ibanez:2012wg}. These authors found a total of 320 models of ADKS class 7506 with a tadpole cancelling hidden sector. 
But, as explained in the beginning of this section, there are two important differences between the method of these authors and ours: 
in \cite{Ibanez:2012wg} only one tadpole solution per local model was considered, and  degeneracies were not removed.
Clearly, it is impossible to make a direct comparison using table of \cite{Ibanez:2012wg}. Instead
we simply identified in our list of 831 solutions those with a  $\mathbb{Z}_N$ symmetry  that does not act on the hidden sector at all. It turns out that this symmetry group is always  $\mathbb{Z}_2$, in agreement with the
results of \cite{Ibanez:2012wg}. The total number is 138, and 32 of these have an additional $\mathbb{Z}_3$ symmetry acting partly on the hidden sector. Some of these 138 models have extra massless $U(1)$'s, but these act
always only on the hidden sector.

Hence considering hidden sectors increases the total number of acceptable models (those without spurious photons) with observable discrete symmetries
from 138 to 372. 
Thus by allowing the discrete symmetry to originate also from hidden $U(1)$ gauge factors one significantly increases the probability of encountering a discrete symmetry in a global compactification. 
This last point is also illustrated in the table \ref{table discrete symmetries -- U(1) factors} where one sees the probability to find a discrete symmetry significantly increases with a larger number of $U(1)$ factors. 
\begin{table}[h]
\centering
{
\begin{tabular}{|c||c||c|c|c|c|c|c|}\hline
\multirow{2}{*}{$\mathbb{Z}_N$'s} &  \multirow{2}{*}{total $\#$ of solutions}  & \multicolumn{6}{|c|}{Number of unitary brane stacks} \\
\cline{3-8}
 & & 3 & 4 & 5 &6 & 7 & 8 \\ \hline \hline
0  &  4189 & 2230 & 885 & 288 & 177 & 127 & 482  \\ \hline
1  &  784 & 37 & 33 & 24 & 155 & 214 & 321    \\ \hline
2  &    45   & & &  & 29 & 4 & 12   \\ \hline
3  &    2   &  & & & &  & 2  \\ \hline
\end{tabular}}
\caption{\small Discrete symmetries in comparison to present $U(1)$ factors.
\label{table discrete symmetries -- U(1) factors} 
}
\end{table}

Let us now display what type of discrete symmetries are realized as well as their frequency (see table \ref{table different discrete symmetries and their frequency}). One observes that $\mathbb{Z}_N$ symmetries with $N\neq 2, 3$ are very rarely realized. This may be related to the fact that the visible sector consists of the SM sector, which only allows for $\mathbb{Z}_2$ and $\mathbb{Z}_3$ discrete symmetries as well as products of it\footnote{Here the SM sector contains also 3 right-handed neutrinos. In the absence of the latter one may also find $\mathbb{Z}_9$ and $\mathbb{Z}_{18}$ symmetries \cite{Dreiner:2005rd,Anastasopoulos:2012zu}}. On the other hand we find also models in which the discrete symmetry lives purely in the hidden sector and nevertheless we find that factors $\mathbb{Z}_p$ with $p$ prime and $p \geq 5$ are not realized.

Again we should stress that our sample is not large enough to draw any serious conclusion, but it is tempting to conclude that abelian discrete $\mathbb{Z}_N$ symmetries with large $N$ are generically rather rarely realized, which also in agreement with previous searches \cite{BerasaluceGonzalez:2011wy,Ibanez:2012wg,Marchesano:2013ega,Honecker:2013hda,Honecker:2013kda}. 
\begin{table}[h] \centering
\begin{tabular}{|c|c|c|c|c|c|}
\hline
$\mathbb{Z}_2$ & $\mathbb{Z}_3 $  &  $\mathbb{Z}_4$ & $\mathbb{Z}_2 \times \mathbb{Z}_2 $ & $\mathbb{Z}_2 \times \mathbb{Z}_3 $ & $\mathbb{Z}_2 \times \mathbb{Z}_2 \times \mathbb{Z}_3 $ \\
\hline \hline
306   & 453 & 23 & 12 & 33 & 2 \\ \hline
 \end{tabular}\nn
\caption{\small {Discrete symmetries in globally consistent models of type Nr. 7506.}} 
\label{table different discrete symmetries and their frequency}
\end{table}

In the following we want to address the question whether in this sample there exist constructions that exhibit one of the intriguing discrete symmetries, such as matter parity, Baryon triality or even proton hexality. Instead of looking at the whole set, we restrict ourselves to the models that have only one massless $U(1)$, namely the hypercharge, in the visible sector. We still consider models that have more than just a single massless $U(1)$ but we require that in that case none of the SM matter fields is charged with respect to the latter. 
From the original sample of 5020 models 4028 global realizations satisfy this constraints. 
Table \ref{table visible discrete symmetries}  reveals that from this sample of 4028 models around 11\% exhibit a discrete symmetry (see also table \ref{table discrete symmetries in total}). 

Matter parity, that forbids all R-parity violating terms, is  realized in around 5\% of all the constructions. 
\begin{table}[h]
\centering
{
\begin{tabular}{|c||c||c|c|c|c|c|c|}\hline
\multirow{2}{*}{$\mathbb{Z}_N$'s} &  \multirow{2}{*}{total $\#$ of solutions}  & \multicolumn{6}{|c|}{Number of unitary brane stacks} \\
\cline{3-8}
 & & 3 & 4 & 5 &6 & 7 & 8 \\ \hline \hline
no $\mathbb{Z}_N$   &  3602 & 2230 & 849     & 261 & 131 & 52 &   79 \\ \hline \hline

hidden $\mathbb{Z}_N$   &  54 &  &      &  &  2 & 17 & 35   \\ \hline \hline

$R_2$  &  227 & 37 & 32 & 7 & 39 & 29 & 83    \\ \hline \hline

$ L_3 R_3 $  &  35    &   &    & 8  & 25 &  & 2   \\ \hline
$R_3 $     &  2      &  & & & & 1  & 1  \\ \hline
$L_3$    &     38 &  & 1 & 7& 19& 11&  \\ \hline 
$ L_3 R^2_3 $ & 36 & & & 2 & 24& 8 & 2     \\ \hline \hline
$ L^2_6 R^5_6$  & 0  & & & & & &   \\ \hline
$R_6$   & 0  & & & & & &   \\ \hline
$L^2_6 R^3_6 $ & 0  & & & & & &    \\ \hline
$L^2_6 R_6 $  & 34  & & & & 28 & 4 &  2\\ \hline
\end{tabular}}
\caption{\small Discrete symmetries in comparison to present $U(1)$ factors.
\label{table visible discrete symmetries} 
}
\end{table}
On the other hand $\mathbb{Z}_3$ symmetries do not show up that frequently. Moreover, while each of the three discrete symmetries $L_3$, $L_3 R^2_3$ and $L_3 R_3$ (baryon triality) appears in one out of 100 models, $R_3$ is significantly suppressed. Finally, the simultaneous appearance of a $\mathbb{Z}_2$ and $\mathbb{Z}_3$, i.e. the presence of a $\mathbb{Z}_6$, is also rather unlikely, and if only appears as $L^2_6 R_6 $, as a product of $R_2$ and $L_3 R^2_3$. Such a $\mathbb{Z}_6$ is however phenomenologically undesired, since it forbids the generation of a mass term for the right-handed neutrinos. 

Table \ref{table visible discrete symmetries} reveals once more that allowing for more $U(1)$ gauge factors enhances significantly the probability of finding a discrete symmetry. For instance there are no $\mathbb{Z}_3$ symmetries for models with just three $U(1)$ (the SM $U(1)$- ) gauge factors. This is in agreement with the findings of \cite{Ibanez:2012wg}, where the authors did not find any $\mathbb{Z}_3$ symmetry in the global embeddings of local Gepner configurations of type Nr. 7506. On the other hand we find all the globally consistent $\mathbf{Z}_2$ ($R_2$ ) realizations also found in \cite{Ibanez:2012wg}, where the matter parity arises as a linear combination of only visible $U(1)$-factors.

The phenomenologically desired $\mathbb{Z}_6$ Proton hexality  $ L^2_6 R^5_6$, the product of matter parity  $R_2$ and baryon triality $L_3 R_3$ does not turn up in any of the models. This may change once one allows for more hidden D-brane stacks. Moreover, here we focus on a particular choice of local SM Gepner configuration. For a different choice there may be very well models that exhibit proton hexality. As shown in \cite{Anastasopoulos:2012zu} different hypercharge embeddings and, moreover, different local D-brane configurations favor different discrete symmetries. It would be interesting to extend the search performed here to other ADKS classes.  

Summarizing, this search revealed that allowing the discrete symmetry to originate also from hidden stacks, increases significantly the probability to find a discrete symmetry in globally consistent MSSM Gepner constructions. While matter parity seems to be favored for models of type Nr. 7506, $\mathbb{Z}_3$ symmetries seem rather suppressed. However, in comparison to the search of  \cite{Ibanez:2012wg}  they $do$ appear. More importantly, we find 35 globally consistent models that exhibit baryon triality, a discrete symmetry that forbids beyond the baryon violating term $U_R D_R D_R$ also the dangerous dimension 5 proton decay operators $Q_L Q_L Q_L L $ and $U_R U_R D_R E_R$. On the other hand we do not find any model that possess proton hexality $ L^2_6 R^5_6$, the holy grail of abelian discrete symmetries.  

\section{Conclusions}
In this paper we investigate the presence of abelian discrete symmetries in globally consistent Gepner constructions, where the abelian discrete symmetries originate from continuous $U(1)$ symmetries that generically become massive via St\"uckelberg couplings symmetries and are preserved on perturbative as well as non-perturbative level in the low energy effective action. More precisely, we extend the work of \cite{Ibanez:2012wg} by allowing the discrete symmetries to be a linear combination of $U(1)$ factors of the visible as well as the hidden sector. 

Our systematic search within globally consistent MSSM-like Gepner constructions reveals that around 17\% of the globally consistent models exhibit a discrete symmetry, whereas in most cases (85\%) the discrete symmetry acts on the visible sector, forbidding some of the undesired couplings. In contrast to the search performed in \cite{Ibanez:2012wg} we find models that exhibit Baryon triality and, moreover, many more MSSM-like Gepner constructions compared to \cite{Ibanez:2012wg} who possess a $\mathbb{Z}_2$ that can be interpreted as matter parity. However, none of the 5028 globally consistent MSSM-like Gepner models exhibits, matter parity and Baryon triality simultaneously, thus none of the models possesses proton hexality. More generally, we encounter all $\mathbb{Z}_2$ and  $\mathbb{Z}_3$ of the MSSM that are compatible with the discrete gauge anomaly constraints are realized, however only one of the possible $\mathbb{Z}_6$, namely $L^2_6 R_6 $ is realized. In summary our analysis shows that allowing the discrete symmetry being a linear combination of $U(1)$ factors not only of the visible sector but also the hidden sector increases the probability of having a discrete symmetry significantly.

In this work we focus on the Madrid hypercharge embedding (see eq. \eqref{eq hypercharge madrid}) and, moreover, on one particular ADKS class, namely on the one with ADKS-label 7506. Within this ADKS class we generate 5028 globally consistent, however distinct, Gepner models that serve as the sample for our search for discrete symmetries. Given that sample we investigate in addition to discrete symmetries also the presence of massless $U(1)$ symmetries in the low energy effective action. Even though the here considered sample is too small to make any general claims, it is tempting to draw the following, also naively expected, conclusion. A larger number of $U(1)$ factors increases the probability of having additional massless $U(1)$'s or discrete symmetries in the low energy effective theory. 

In this context a multiverse/landscape question naturally arises, namely why is a single $U(1)$ and a discrete symmetry singled out from the sample, other than by mere phenomenological constraints?
We can only speculate about this. Our statistics is too limited, and furthermore this is not genuine landscape statistics based on moduli-stabilized non-supersymmetric deSitter vacua. But the first thing we can say is that the suppression of the desirable features within our set is not by huge factors. Even  though we did not find the most desirable discrete symmetry, proton hexality, within our sample of 5020 tadpole solutions, the presence of other
discrete symmetries suggests suppression factors of order 10 to 1000.
Furthermore several of the observed features are potentially anthropically favored. 
This is certainly true for the absence of dimension-4 proton decay. It may also be true for the absence of additional $U(1)$'s coupling to quarks and leptons: such $U(1)$'s lead to additional Coulomb forces radically altering atomic and nuclear physics. On the other hand, proton decay by dimension five operators in supersymmetric models would be merely observable, but not fatal, and is a more serious landscape issue.  But given the uncertainty in the statistics this is not yet extremely worrisome, especially not in comparison to other landscape issues. See also \cite{Dine:2005gz} for a discussion of discrete symmetries in the landscape from a different perspective.

It would be desirable to extend the analysis to other ADKS classes with the same hypercharge embedding, but also to different hypercharge embeddings. In particular it is interesting to see whether the findings of \cite{Anastasopoulos:2012zu}, that in local D-brane configurations different hypercharge embeddings favor different discrete symmetries holds also true in the global setting. Moreover, it is interesting to see whether global embeddings of other ADKS classes allow for family-depended abelian discrete symmetries with  an intriguing low energy behavior that go beyond the ones discussed in \cite{Dreiner:2005rd}. We leave this for future work.  Finally, it would be interesting to study the presence of discrete symmetries in generic globally consistent string models, models that typically do not possess a SM sector. 

\section*{Acknowledgements}

We thank P. Soler  for useful discussions. P.A. is supported by the Austrian Science Fund (FWF) program M 1428-N27.
R.R. was partly supported by the German Science Foundation (DFG) under the Collaborative Research Center (SFB) 676 ``Particles, Strings and the Early Universe".
R.R. acknowledges the support from the FP7 Marie Curie Actions of the European Commission, via the Intra-European Fellowships (project number: 328170). P.A. and R.R. would like to thank CERN for hospitality during parts of this work.
This work has been partially supported by funding of the Spanish Ministerio de Econom\'ia y Competitividad, Research Project FIS2012-38816, and by the Project CONSOLIDER-INGENIO 2010, Programme CPAN (CSD2007-00042).

\newpage
\bibliographystyle{utphys}

\providecommand{\href}[2]{#2}\begingroup\raggedright\endgroup

\end{document}